# Measures of operator entanglement of two-qubit gates

S. Balakrishnan and R. Sankaranarayanan
*Department of Physics, National Institute of Technology, Tiruchirappalli 620015, India*

Two different measures of operator entanglement of two-qubit gates, namely, Schmidt strength and linear entropy, are studied. While these measures are shown to have one-to-one relation between them for Schmidt number 2 class of gates, no such relation exists for Schmidt number 4 class, implying that the measures are inequivalent in general. Further, we establish a simple relation between linear entropy and local invariants of two-qubit gates. The implication of the relation is discussed.

## I. INTRODUCTION

As entangled states offer a wide variety of applications in teleportation, cryptography, computation and quantum game theory, much of the theoretical studies on quantum information processing revolve around the entanglement characterization [1]. Equally, nonlocal two-qubit operators (gates) are also investigated in detail as they are capable of producing entanglement when acting on a state. A few characterizing tools have been developed to understand the nonlocal features of quantum gates. One such tool is entangling power, which quantifies the average entanglement produced by a gate when acting on product states [2, 3]. The concept of entangling power has been studied with various motivations [4-6].

Nonlocal attributes of two-qubit gates are uniquely associated with local invariants, which are unaffected by local operations [7]. From the canonical decomposition, a two-qubit gate is uniquely represented by three geometrical points as well. By connecting local invariants and geometrical points of a two-qubit gate, it is found that *all* nonlocal gates form an irreducible geometry of tetrahedron (known as Weyl chamber). It is also known that exactly half of the nonlocal gates are capable of producing maximally entangled state when acting on some separable states, and they are called as perfect entanglers. Geometrically, the perfect entanglers form a polyhedron within the Weyl chamber [8].

On other hand, a quantum operator itself may be represented in Hilbert-Schmidt space. For a two-qubit gate, this space is four dimensional. From this perspective, entanglement of an operator is also relevant to quantify the nonlocal attributes. While entanglement production of an operator is widely studied, the entanglement of an operator is addressed in a limited way. In this work, we consider two-qubit gates which are represented in an operator-Schmidt decomposition form. In this representation, the number of nonzero (Schmidt) coefficients of a gate is called as Schmidt number. It is known that local gates have Schmidt number 1 and nonlocal gates have Schmidt number 2 or 4 [9]. Entanglement of an operator is usually quantified by two measures, namely, Schmidt strength $K_{Sch}(U)$ [9] and linear entropy $L(U)$ [2, 3]. We shall note that the linear entropy of an operator is related, though not in a simple way, to the entangling power [2, 3].

In this paper, we attempted to check if the above two measures of operator entanglement are equivalent. We find that there exists a one-to-one relation between $K_{Sch}(U)$ and $L(U)$ for Schmidt number 2 class of gates. However, it is shown numerically that no such relation between the measures seem to exist for Schmidt number 4 class, implying that $K_{Sch}(U)$ and $L(U)$ are *inequivalent* measures of operator entanglement. Expressing linear entropy in terms of geometrical points, we find the gates having maximum linear entropy, and they lie on one edge of the Weyl chamber. Further, we obtained a simple relation between linear entropy and local invariants, signifying the role of local invariants in measuring the operator entanglement.

## II. PRELIMINARIES
### A. Geometry

An arbitrary two-qubit gate $U \in$ SU(4) can be written in the following form, which is known as canonical decomposition [10]:

$$U = k_1 exp\left\{\frac{i}{2}\left(c_1 \sigma_x^1 \sigma_x^2 + c_2 \sigma_y^1 \sigma_y^2 + c_3 \sigma_z^1 \sigma_z^2\right)\right\} k_2 \tag{1}$$

where $\sigma_x, \sigma_y, \sigma_z$ are Pauli matrices and $k_1, k_2 \in$ SU(2)⊗SU(2). Two unitary operators $U, U_1 \in$ SU(4) are called locally equivalent if they differ only by local operations: $U = k_1 U_1 k_2$. A class of gates differ from $U$ only by local operations is referred as local equivalence class $[U]$. Makhlin introduced the notion of local invariants which *uniquely* characterize the local equivalence class [7, 8]. Local invariants $G_1, G_2$ and a point $[c_1, c_2, c_3]$ corresponding to the gate $U$ are related as [8]

$$G_1 = cos^2 c_1 cos^2 c_2 cos^2 c_3 - sin^2 c_1 sin^2 c_2 sin^2 c_3 + \frac{i}{4} sin 2c_1 sin 2c_2 sin 2c_3, \tag{2a}$$

$$G_2 = 4cos^2 c_1 cos^2 c_2 cos^2 c_3 - 4sin^2 c_1 sin^2 c_2 sin^2 c_3 - cos 2c_1 cos 2c_2 cos 2c_3. \tag{2b}$$

From this relation, for given local invariants $(G_1, G_2)$, the point $[c_1, c_2, c_3]$ in a 3-torus geometry (with period $\pi$) is identified. In other words, the gate $U$ or its equivalence class $[U]$ is uniquely characterized by the point $[c_1, c_2, c_3]$ as well. The symmetry reduced 3-torus takes the form of tetrahedron (the Weyl chamber). A two-qubit gate is called a perfect entangler if it produces a maximally entangled state when acting on some separable input state [7, 8]. Perfect entanglers constitute a polyhedron within the Weyl chamber, as shown in Fig. 1.

An alternate representation of two-qubit gate $U$ is the operator-Schmidt decomposition

$$U = \sum_l s_l\, A_l \otimes B_l \tag{3}$$

where $s_l \geq 0$ are called as Schmidt coefficients and $A_l(B_l)$ are orthonormal operator bases for system $A(B)$ [9]. In this representation, the number of nonzero Schmidt coefficients of an operator is defined as *Schmidt number*. It is known that local gates have Schmidt number 1 and nonlocal gates have Schmidt number 2 or 4 [9]. Using the geometrical points of the edges, Schmidt coefficients of six edges of tetrahedron (Weyl chamber) and nine edges of polyhedron are already computed [11].

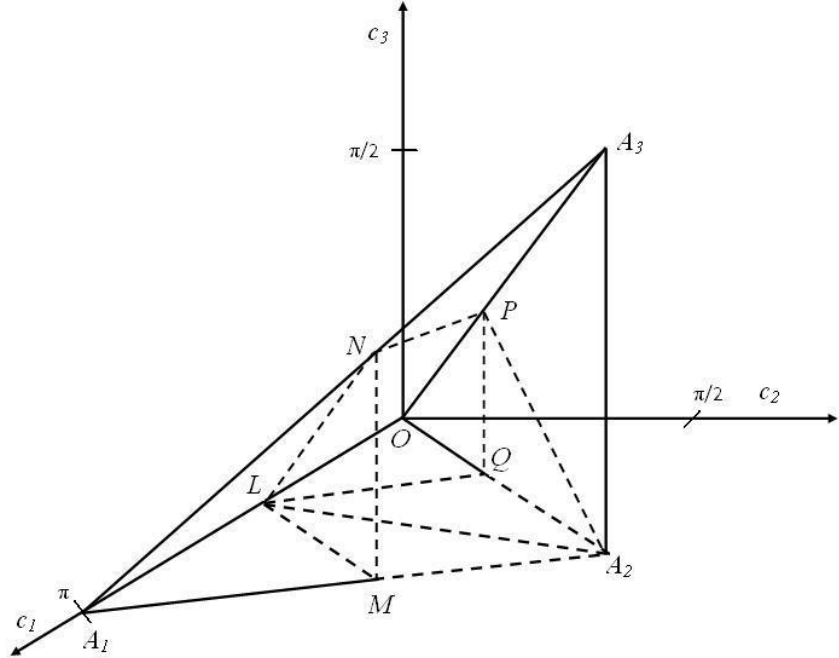

FIG. 1. Tetrahedron $OA_1A_2A_3$ (Weyl chamber) is the geometrical representation of nonlocal two-qubit gates. Polyhedron $LMNPQA_2$ (shown in dashed lines) corresponds to the perfect entanglers. The points $L$, $M$, $N$, $P$, and $Q$ are midpoints of the tetrahedron edges $OA_1$, $A_2A_1$, $A_1A_3$, $OA_3$, and $OA_2$ respectively. The points $L = [\pi/2, 0, 0]$, $A_2 = [\pi/2, \pi/2, 0]$ and $A_3 = [\pi/2, \pi/2, \pi/2]$ correspond to *CNOT*, *Double-CNOT* and *SWAP* gates respectively. The origin represents the local gate. The line $LA_2$ represents special perfect entanglers.

## B. Operator entanglement

Here, we define various measures of operator entanglement. Since the Schmidt coefficients satisfy $\sum_{l=1}^{4} s_l^2 = 1$, $s_l^2$ form a probability distribution. Exploiting this property, Schmidt strength (a measure of operator entanglement) is defined as the Shannon entropy of the distribution $s_l^2$,

$$K_{Sch}(U) = -\sum_{l=1}^{4} s_l^2 \log_2 s_l^2, \tag{4}$$

such that $0 \leq K_{Sch}(U) \leq 2$ [9]. While $K_{Sch}(U) = 0$ for local gates, $K_{Sch}(U) = 2$ for the edge $A_2A_3$ [11]. In other words, local gates are not entangled and the gates on the edge $A_2A_3$ are maximally entangled.

Following Ref. [2], we define linear entropy of an operator $U$,

$$L(U) = 1 - \frac{1}{d^4} \langle U^{\otimes 2}, T_{1,3} U^{\otimes 2} T_{1,3} \rangle, \tag{5}$$

where $\langle A, B \rangle = tr(A^\dagger B)$ is referred to as Hilbert-Schmidt scalar product and $T_{1,3}$ is the permutation operator defined as $T_{1,3}|a,b,c,d\rangle = |c,b,a,d\rangle$ on a four-qubit system. For two-qubit gates, the above expression can be rewritten as

$$L(U) = 1 - \frac{1}{16} tr(U^{\dagger \otimes 2} T_{1,3} U^{\otimes 2} T_{1,3}). \tag{6}$$

In terms of Schmidt coefficients, the linear entropy is defined as

$$L(U) = 1 - \sum_{l=1}^{4} s_l^4, \tag{7}$$

such that $0 \leq L(U) \leq 3/4$. While $L(U) = 0$ for local gates, $L(U) = 3/4$ for the edge $A_2A_3$. That is, the gates lie on the edge $A_2A_3$ are maximally entangled in this measure as well. Some well known gates lying on the edge $A_2A_3$ are $SWAP\ [\pi/2, \pi/2, \pi/2]$ and $DCNOT\ [\pi/2, \pi/2, 0]$. However, it is not known if any gates other than the edge $A_2A_3$ are also maximally entangled. Another measure of operator entanglement is concurrence $C(U)$, which is defined only for Schmidt number 2 gates in the following way [12]:

$$C(U) = 2 s_1 s_2. \tag{8}$$

One can show that

$$C(U) = \sqrt{2L(U)}, \tag{9}$$

implying the one-to-one relation between the two measures for Schmidt number 2 class of gates.

## III. SCHMIDT STRENGTH AND LINEAR ENTROPY

Having defined various measures of operator entanglement, in this section we dwell on Schmidt strength $K_{Sch}(U)$ and linear entropy $L(U)$ for a variety of gates. It is known that controlled unitary gates (the edge $OA_1$ in the Weyl chamber) correspond to Schmidt number 2 class and all other gates correspond to Schmidt number 4 [11]. In what follows, we obtain a simple relation between $K_{Sch}(U)$ and $L(U)$ for Schmidt number 2 gates. Since $s_1^2 + s_2^2 = 1$, $L(U) = 1 - (s_1^4 + s_2^4) = 2s_1^2 s_2^2$. Since $s_1^2$ satisfies the relation $s_1^4 - s_1^2 + L(U)/2 = 0$, we have

$$s_1^2 = \frac{1 \pm \sqrt{1 - 2L(U)}}{2} \quad \text{and} \quad s_2^2 = \frac{1 \mp \sqrt{1 - 2L(U)}}{2}.$$

Substituting the above expressions for Schmidt coefficients in Eq. (4), we have

$$K_{Sch}(U) = -\left[\frac{1 \pm \sqrt{1-2L(U)}}{2} \log_2 \frac{1 \pm \sqrt{1-2L(U)}}{2} + \frac{1 \mp \sqrt{1-2L(U)}}{2} \log_2 \frac{1 \mp \sqrt{1-2L(U)}}{2}\right]$$

(10)

and the same is plotted in Fig. 2, showing the one-to-one relation between the two measures. Further, we observe from Eq. (9) and (10) that all the three measures of operator entanglement, namely, Schmidt strength, linear entropy, and concurrence, are equivalent for Schmidt number 2 gates.

Since the relation between the above measures for Schmidt number 4 class of gates is nontrivial, here we compute these measures for all the geometrical edges of two-qubit gates whose Schmidt coefficients are already known [11]. In what follows, $K_{Sch}(U)$ versus $L(U)$ are plotted for the four edges of Weyl chamber [the edge $OA_1$ corresponds to controlled unitary gates; for the edge $A_2A_3$, $K_{Sch}(U) = 2$ and $L(U) = 3/4$] and nine edges of polyhedron. Figure 2 shows the plot for the edges of Weyl chamber, from which we observe that there exists a one-to-one relation between the two measures for each edge.

Plot of $K_{Sch}(U)$ versus $L(U)$ for nine edges of the polyhedron is shown in Fig. 3. Since the polyhedron edges *LQ* and *LM* are locally equivalent to each other, they possess same form of $K_{Sch}(L(U))$. Similarly, the edges $A_2M$ and $A_2Q$ are also locally equivalent and they assume the same form of $K_{Sch}(L(U))$. Interestingly, the measures have same form of $K_{Sch}(L(U))$ for the edges *QP* and *MN* as well, though the edges are not locally equivalent to each other. This can be understood partially in terms of local invariants of the edges as discussed in the following section. Here also we find that there exists a one-to-one relation between the two measures for each edge of the polyhedron. Nevertheless, for the edge *LQ*, as $L(U)$ increases $K_{Sch}(U)$ decreases.

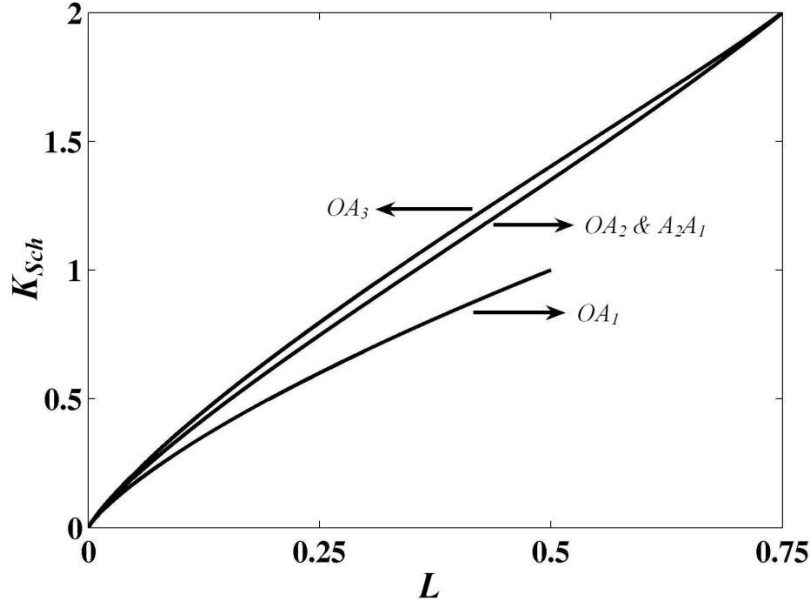

FIG. 2. $K_{Sch}(U)$ Vs. $L(U)$ for the Weyl chamber edges.

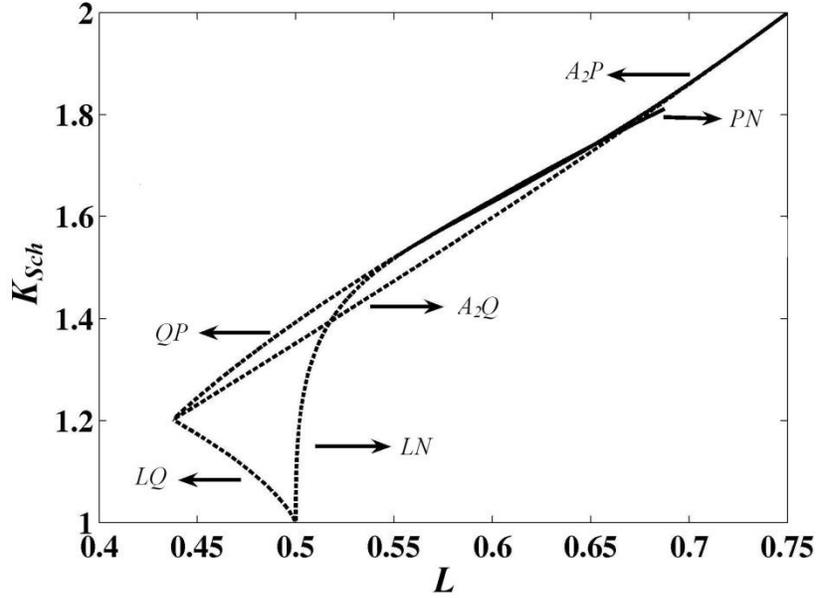

FIG. 3. $K_{Sch}(U)$ Vs. $L(U)$ for the polyhedron edges.

In order to analyze the general behavior of the measures, $K_{Sch}(U)$ and $L(U)$ are computed for arbitrary two-qubit gates distributed uniformly in the geometrical space, and they are plotted in Fig. 4. It is clear from the plot that, in general, the Schmidt strength and linear entropy do not possess a one-to-one relation between them. Further, the correlation function between the measures is 0.0705. The two Weyl chamber edges $OA_1$, $OA_3$ and special perfect entanglers, $LA_2$ form the boundaries in the $K_{Sch} - L$ plane of two-qubit gates. This shows that, in general, there exists no simple relation between the two measures of operator entanglement.

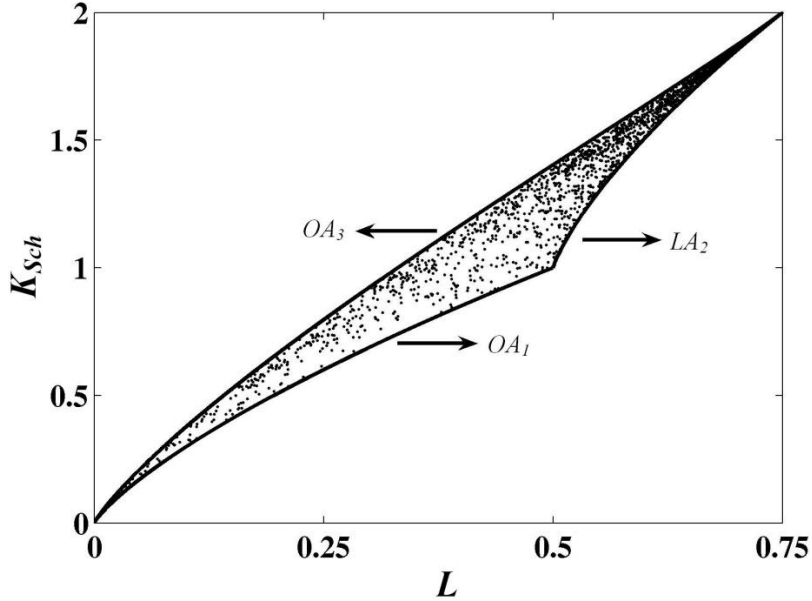

FIG. 4. $K_{Sch}(U)$ Vs. $L(U)$ for arbitrary two-qubit gates.

## IV. LINEAR ENTROPY AND LOCAL INVARIANTS

As linear entropy is easily computable and is related to the entangling power, in this section we study the linear entropy of two-qubit gates in detail. Considering the nonlocal part of two-qubit gate $U$ as given in Eq. (1), from Eq. (6) the linear entropy can be expressed as

$$L(U) = 1 - \frac{1}{4}(1 + \cos^2 c_1 \cos^2 c_2 + \cos^2 c_2 \cos^2 c_3 + \cos^2 c_3 \cos^2 c_1). \tag{11}$$

Here we have an expression for operator entanglement in terms of the geometrical points $[c_1, c_2, c_3]$, which is useful for the following analysis. From Eq. (11), it is easy to see that $L(U) = 3/4$, the maximum value, only for $c_1 = c_2 = \pi/2$ with arbitrary $c_3$. In other words, the gates $\left[\frac{\pi}{2}, \frac{\pi}{2}, \varphi\right]$, where $0 \leq \varphi \leq \pi/2$ constitute the edge $A_2A_3$ for which linear entropy is maximum. It implies that the gates lying *only* on the edge $A_2A_3$ are maximally entangled and no other two-qubit gates are maximally entangled. Similarly, $L(U) = 0$ only for $c_1 = c_2 = c_3 = 0$, which refers to the local gate.

After some algebraic simplifications, the above expression can also be written as

$$L(U) = 1 - \frac{1}{8}\big(3 + 2|G_1(U)| + G_2(U)\big). \tag{12}$$

Thus we arrive at a simple relation between linear entropy and the local invariants. From this relation, it is clear that the gates having the same $|G_1|$ and $G_2$ must necessarily possess the *same* linear entropy, implying that the gates are equally entangled. For example, the edges *QP* and *MN* of polyhedron are such that their $|G_1|$ and $G_2$ are the same [13], resulting in the same linear entropy. Further, the above expression also facilitates to understand the operator entanglement of perfect entanglers. It is known from our earlier studies that the local invariants of perfect entanglers are such that $0 \leq |G_1| \leq 1/4$ and $-1 \leq G_2 \leq 1$ [14]. In other words, the linear entropy of perfect entanglers are such that $0.4375 \leq L(U) \leq 0.75$, where the minimum and maximum values correspond to the gates $Q$ and $A_2$, respectively (see Fig. 1). Therefore, *DCNOT* is the only perfect entangler which is maximally entangled.

Further, the linear entropy of an operator $US$, where $S$ is *SWAP*, is found to be

$$L(US) = 1 - \frac{1}{8}\big(3 + 2|G_1(U)| - G_2(U)\big). \tag{13}$$

The entangling power of an operator $e_p(U)$ is defined as the average entanglement produced when it acts on all possible input product states distributed uniformly in the state space [2]. The entangling power is related to the linear entropy as [3, 12]

$$e_P(U) = \frac{4}{9}[L(U) + L(US) - L(S)], \tag{14}$$

where $L(S) = 3/4$. By substituting Eq. (12) and Eq. (13) in the above equation, we have

$$e_P(U) = \frac{2}{9}[1 - |G_1(U)|], \tag{15}$$

which is already obtained in our earlier study on entangling power [14]. The last relation implies that the gates with the same $|G_1(U)|$ possess the same entangling power. From Eqs. (12) and (15), the local invariants associated to a gate are found to quantify the average entanglement produced and the operator entanglement. Thus, the local invariants are related to two different nonlocal attributes of an operator, namely, the entanglement production and operator entanglement.

## V. CONCLUSION

In this paper, we study two different operator entanglement measures of two-qubit gates, namely, Schmidt strength and linear entropy. For Schmidt number 2 gates, we have established a simple one-to-one relation between the measures. It is known for Schmidt number 2 gates that entangling power is proportional to the linear entropy [4], which is shown to possess one-to-one relation with Schmidt strength. Hence, the entanglement production and operator entanglement are proportional only for Schmidt number 2 class of gates. In other words, if a Schmidt number 2 class of gate is more entangled, it produces more entanglement on the states. However, our numerical calculation indicates that such a relation between the measures does not exist for Schmidt number 4 gates. This implies that, in general, Schmidt strength and linear entropy are *not* equivalent measures of operator entanglement.

Further, we have studied linear entropy in more detail as it is related, though not in a simple way, to the entangling power of a gate. In particular, we are able to express linear entropy in terms of geometrical points of a gate. From the expression, we identify the gates having a maximum linear entropy and they lie on one geometrical edge ($A_2 A_3$) of the Weyl chamber. In other words, all the maximally entangled gates lie on the edge $A_2 A_3$ of the Weyl chamber. The above expression also facilitates to obtain a simple relation between linear entropy and local invariants. The relation implies that gates having the same $|G_1|$ and $G_2$ must necessarily possess the same linear entropy. From the relation, we deduce for perfect entanglers that $0.4375 \leq L(U) \leq 3/4$, with $Q$ (see Fig.1) and $DCNOT$ possessing minimum and maximum entropy respectively. Further, the relation between linear entropy and local invariants suggests that linear entropy is more useful than the Schmidt strength in quantifying the nonlocal attributes of an operator.

It is known that entanglement production of a gate, as measured by the entangling power, and operator entanglement, as measured by linear entropy, are different approaches in charactering nonlocal attributes of a gate. In this work, the two approaches are shown to be related to the local invariants. Thus the local invariants play a central role in the investigations of nonlocal two-qubit gates.

## ACKNOWLEDGMENT

S. B acknowledges financial support from the Council of Scientific and Industrial Research, New Delhi, India.

___________________________________________________________________